\documentstyle[aps,preprint]{revtex}
\input{psfig}

\def \vs {{\bf v_s}}

\def \d  {{\rm d}}

\begin{document}

\bibliographystyle{unsrt}
\newcommand{\be}{\begin{equation}}
\newcommand{\ee}{\end{equation}}
\newcommand{\bea}{\begin{eqnarray}}
\newcommand{\eea}{\end{eqnarray}}


\title
{Combined Paramagnetic and Diamagnetic Response of YBCO}

\author{N. D. Whelan and J. P. Carbotte}
\address{Department of Physics and Astronomy, McMaster University,
Hamilton, Ontario, Canada L8S~4M1}

\date{\today}

\maketitle

\begin{abstract}
It has been predicted that the zero frequency density of states of
YBCO in the superconducting phase can display interesting anisotropy
effects when a magnetic field is applied parallel to the copper-oxide
planes, due to the diamagnetic response of the quasi-particles. In
this paper we incorporate paramagnetism into the theory and show that
it lessens the anisotropy and can even eliminate it altogether. At the
same time paramagnetism also changes the scaling with the square root
of the magnetic field first deduced by Volovik leading to an
experimentally testable prediction.  We also map out the analytic
structure of the zero frequency density of states as a function of the
diamagnetic and paramagnetic energies. At certain critical magnetic
field values we predict kinks as we vary the magnetic field. However
these probably lie beyond currently accessible field strengths.

\end{abstract}



\section{Introduction}

\noindent
In 1993 Volovik \cite{volovik} predicted that the density of states of
a $d$-wave superconductor is dominated by delocalised nodal
quasi-particles and scales as the square root of the magnetic
field. Since then there has been a lot of research on the role of
nodal quasi-particles in determining the electronic density of states
of the cuprate superconductors. Experimentally this is probed by
looking at the specific heat as a function of magnetic field
\cite{experiments1,experiments2,experiments3,salamon}. This is a
bulk property and therefore complements STM and ARPES measurements
which are surface probes. (On the other hand, it is a somewhat less
direct probe of the electronic density of states.) Theoretically,
groups have studied the effect of dirty superconductors
\cite{kubhirsh}, the role of the copper oxide chains in affecting the
specific heat \cite{us} and the anisotropic dependences on magnetic
field direction \cite{vekhter}. In particular, the last paper
predicted that the observed dependence of the density of states of an
in plane magnetic field would depend on its orientation relative to
the gap nodes. That paper only accounted for the diamagnetic response
of the quasi-particles, as in Volovik's original paper. In this paper
we generalise this to include the paramagnetic response of the
quasi-particles. The quasi-particle paramagnetism has been considered
in different contexts in \cite{para1,para2}. We focus on YBCO because only
it has sufficient transport in the $c$ direction to allow the usual
London theory to apply so that vortices to form out of the CuO planes
\cite{vekhter}. Other relevant papers that treat different aspects of
the problem include \cite{barash,franz,melnikov}.

In the following section we present a general discussion of the theory
and define the various terms. In section \ref{energy_scales} we
discuss the energy scales specific to YBCO. The two subsequent
sections present the results for the physical case of relatively small
magnetic field for the field perpendicular and parallel to the CuO
planes, respectively.  With small fields, a particularly simple
approximation can be invoked leading to analytic results. In the
parallel case, we predict a loss of anisotropy at sufficiently large
magnetic fields. In section \ref{heuristic} we present a more physical
argument for the loss of anisotropy. We summarise our results in the
conclusion. In the appendix, we allow for fields of arbitrary strength
but only in the situation where the field is perpendicular to the
copper-oxide planes.

\section{General Theory}

\noindent
The approach we use is to consider a BCS-like formalism \cite{mahan}
with a $d$-wave gap. Since YBCO is a type-II superconductor, there are
vortices above a critical field $H_{c1}$ and we will focus on the
regime $H_{c1}\ll H\ll H_{c2}$. This condition ensures that $\xi\ll
R\ll\lambda$ where $\xi$, $R$ and $\lambda$ are the coherence length,
vortex radius and penetration depth, respectively. In fact, $H_{c1}$
is so small ($\sim 100$ Gauss \cite{ginsberg}) and $H_{c2}$ so large
($\sim 100$ Tesla) that one is always in this experimental range.
Around each vortex there is a rotating superfluid condensate.
Quasiparticles are excited out of this condensate and are accordingly
Doppler-shifted by the superfluid velocity. The effect of this Doppler
shift is to give a finite density of states, even at zero frequency,
and we refer to this as the diamagnetic effect. In addition, the
individual quasi-particles have a paramagnetic response to the
magnetic field due to their intrinsic spin and we refer to this as the
paramagnetic effect. In this paper, we are interested in the
competition between these two effects, there being two important
geometries. The simpler one is when the field is parallel to the
$c$-axis ($H \parallel c$) in which event we can study effects for all
magnetic field values. The more physically interesting geometry,
however, is that the field is parallel to the $a-b$ plane ($H\parallel
ab$). In this case the diamagnetism results in an anisotropic
dependence of the density of states on the direction of the applied
magnetic field, with a greater value when the field points along the
gap antinodal direction and a lesser value when the field points along
the nodal direction. We show that the paramagnetism lessens this
effect and at a critical magnetic field makes it disappear altogether.

Due to the paramagnetic shift, the density of states is broken into
two components which can be written semiclassically as
\be \label{firstequation}
N_0^{(\pm)} \propto {1\over\pi R^2}\int\d{\bf r}\int\d{\bf k} |V|
\delta\left(\left(\sqrt{\xi_{\bf k}^2 + \Delta_{\bf k}^2} \pm \mu
H\right)^2 - V^2\right).
\ee
This is a simple generalisation of the formalism of \cite{vekhter}
where we have included the paramagnetic term $\pm\mu H$. We describe
each factor in turn. The $d$-wave gap is given by $\Delta_{\bf
k}=\Delta_0\cos^2(2\phi)$ and $\xi_{\bf k} = k^2-\mu_c$ so that
$\sqrt{\xi_{\bf k}^2 + \Delta_{\bf k}^2} \pm \mu H$ is the
quasi-particle excitation energy. (Here we have defined the
quasi-particle energy relative to the chemical potential $\mu_c$.) The
integral over ${\bf k}$ (with polar coordinates $k$ and $\phi$) is a
standard trace integral used to determine the density of states. We
will use the delta function to do the radial $k$ integral. The
integral over ${\bf r}$ is an average over one unit cell of the vortex
lattice; $r$ is the distance from the vortex core and $\beta$ is the
vortex winding angle. It is convenient to nondimensionalise the radial
integral to $\rho=r/R$, where $R$ is the inter-vortex distance
\be \label{intvortdist}
R = a^{-1}\sqrt{\Phi_0\over\pi H},
\ee
$a$ is a geometrical constant accounting for the mismatch between the
circular vortices and the regular lattice they fill out and
$\Phi_0=hc/2e$ is a flux quantum.

The factor $V=\vs\cdot{\bf k_F}$ is the Doppler shift which depends on
the condensate velocity $\vs$ (which in turn depends on ${\bf r}$) and
on the Fermi momentum ${\bf k_F}$. We first consider $H \parallel c$;
far from the vortex core (but still within the penetration depth
scale) this is approximately
\be \label{V||c}
V \approx
{E_H\over\rho}\sin(\beta-\phi)
\phantom{\sin\beta\sin(\phi-\alpha)} H \parallel c.
\ee
$E_H=v_Fa\sqrt{\pi H/4\Phi_0}$ is an energy scale associated with the
diamagnetism (and $v_F$ is the Fermi velocity.) This yields
$E_H=2.4\sqrt{H}$ \cite{vekhter} where $E_H$ is measured in $m$eV and
$H$ in Tesla. In practise, when integrating over the winding angle
$\beta$, we change variables to $\beta-\phi$ so the sinusoid is
$\sin\beta$. There are two difference when $H \parallel ab$. Firstly,
the differently aligned vortices implies
\be
\nonumber\\
V \approx {E_H\over\rho}\sin\beta\sin(\phi-\alpha)
\phantom{\sin(\beta-\phi)} {\rm H\parallel ab}.
\ee
There is an additional parameter $\alpha$ which describes the
orientation of the field with respect to the axis directions in the
$ab$ plane. Specifically, when $\alpha=0$, the field points along the
antinodal direction of the gap function while when $\alpha=\pi/4$, the
field points along the nodal direction. The second difference is that
the energy scale $E_H$ is smaller. This is due to the fact that the
vortices must form out of the copper oxide planes. This leads to a
rescaling of the effective Fermi velocity by
$\sqrt{\lambda_{ab}/\lambda_c}$ where $\lambda_i$ is the penetration
depth in direction $i$. $E_H$ is accordingly reduced by a factor of
about 2.5 relative to the $H \parallel c$ case.

For now, we shall keep $V$ arbitrary and thereby include both field
alignments with the same analysis.  To begin, we need to find the
values of $k$ for which the argument of the delta function is
zero. For the $+\mu H$ term this is given by
\be\label{1poss}
\sqrt{\xi_{\bf k}^2 + \Delta_{\bf k}^2} + \mu H - |V| = 0.
\ee
A solution may or may not exist depending on the values of $\phi$,
$\beta$ and $\rho$. (In principle we should also allow for the
solution with $+|V|$, but in practise there are no solutions to that
equation.) We shall refer to the contribution arising from this term
as $N_0^{(+-)}$ (the reason for this name is made clear below.) For
the $-\mu H$ term the analysis is a little more complicated.
Specifically, we now must allow for the possibility of two distinct
roots of the form
\be\label{2poss}
\sqrt{\xi_{\bf k}^2 + \Delta_{\bf k}^2} - \mu H \pm |V| = 0.
\ee
Here we allow for both signs of $|V|$. One should not confuse the
$\pm$ in (\ref{2poss}) with the $\pm$ in the choice of spin direction
in (\ref{firstequation}). We shall call the respective contributions
$N_0^{(-+)}$ and $N_0^{(--)}$ where the first sign refers to $\mu H$
and the second to $|V|$.

As stated, we use the $\delta$ function to do the $k$ integral. This
leads to the following expressions for the three terms (where we
normalise by recalling that for large fields, the total density of
states should approach that of the normal state $\overline{N}$):
\bea \label{general}
{N_0^{(+-)}\over\overline{N}} & = &
{1\over4\pi^2}\int_0^1\d\rho\rho\int_0^{2\pi}\d\beta\int_0^{2\pi}\d\phi
\;P\left({|V|-\mu H\over\sqrt{(|V|-\mu H)^2-{\Delta_{\bf k}^2}}}\right)
\nonumber\\
{N_0^{(--)}\over\overline{N}} & = &
{1\over4\pi^2}\int_0^1\d\rho\rho\int_0^{2\pi}\d\beta\int_0^{2\pi}\d\phi
\;P\left({|V|+\mu H\over\sqrt{(|V|+\mu H)^2-{\Delta_{\bf k}^2}}}\right)
\nonumber\\
{N_0^{(-+)}\over\overline{N}} & = &
{1\over4\pi^2}\int_0^1\d\rho\rho\int_0^{2\pi}\d\beta\int_0^{2\pi}\d\phi
\;P\left({\mu H-|V|\over\sqrt{(\mu H-|V|)^2-{\Delta_{\bf k}^2}}}\right).
\eea
The function $P(x)$ equals $x$ if $x$ is real and positive, otherwise
it is zero (this yields the ranges over which the respective solutions
of Eqs.(\ref{1poss}) and (\ref{2poss}) exist.) Our goal is to do these
three integrals and thereby calculate the density of states and
specific heat. It is convenient at this point to introduce the
dimensionless quantities $\nu=E_H/\Delta_0$ and $\sigma=\mu
H/\Delta_0$ describing the energies associated with the diamagnetism
and paramagnetism respectively.  Mathematically we can treat them as
independent quantities. In reality they are dependent since
$\nu\propto\sqrt{H}$ while $\sigma\propto H$ so that
$\sigma\propto\nu^2$, but this can be imposed later.

We begin by assuming that the fields are small enough to use the
``nodal approximation'', which amounts to replacing the gap function
$\Delta_{\bf k}$ by a local linear approximation in the neighbourhood
of the gap nodes. This leads to a very simple $\phi$ integral. This
approximation is valid in the limit of small $\nu$ and $\sigma$ since
in that event the range of contributing $\phi$ values in
(\ref{general}) is very small and a linear approximation is valid
throughout the allowed integration range. It is convenient to first
take $H\parallel c$. This then sets up the small field $H\parallel ab$
calculation. The small field limit underlying the nodal approximation
is physically relevant since it corresponds to currently accessible
magnetic field strengths. However, it is also interesting to explore
the structure for arbitrary fields; this is done in the appendix.

\section{Energy Scales}\label{energy_scales}

\noindent
We now discuss the energy scales in the problem. It will be convenient
to parameterise the diamagnetic and paramagnetic energies as
follows. We define the coefficients $a$ and $b$ such that
$\nu=a\sqrt{H}$ and $\sigma=bH$ and the functional relationship
between $\nu$ and $\sigma$ is $\sigma=b\nu^2/a^2$. Knowing the
coefficients $a$ and $b$ gives us the full information about the
thermodynamic response of the system. For example, the cross-over
field $H_e$ where $\nu=\sigma$ is given as $(a/b)^2$.

We begin by noting that for YBCO $\Delta_0$ is about 20 $meV$. The
paramagnetic energy is simple since it is just $\mu H$ which equals
$0.058H$ where $H$ is measured in Tesla and the energy in $meV$
(henceforth these units will be understood.) Together these imply
$b=2.9{\rm x}10^{-3}$. The diamagnetic energy scale is trickier; to
date it has been found directly from experimental measurements. This
point is discussed in Refs.~\cite{energy_scales} where it is argued
that $E_H$ is different for $H\parallel c$ and $H\parallel ab$. In the
first case, $E_H \approx 2.4\sqrt{H}$ so that $a=0.12$ and
$\sigma=0.20\nu^2$ . The cross-over field is therefore $H_e=1.7{\rm
x}10^3$T. This is enormous and we can confidently state that for all
experimentally accessible fields, the diamagnetism completely
dominates the paramagnetism.

The $H\parallel ab$ case is more ambiguous. Based on the ratio of the
penetration depths in the $c$ direction and the $a$ and $b$
directions, the authors of Refs.~\cite{energy_scales} estimated that
$E_H\approx 0.7\sqrt{H}$ leading to $a=0.035$ so that the cross-over
field is $H_e=147T$. On the other hand, experiments which compared the
scaling of the specific heat with $\sqrt{H}$ in the $c$ and $a-b$
directions led to the conclusion that $E_H<0.4\sqrt{H}$
\cite{junod}. Taking the upper limit of the inequality for
concreteness, we would have $a=0.020$ so that the cross-over field is
$H_e=48T$. While the cross-over happens at rather large fields, we
will show that the loss of anisotropy is apparent at smaller fields 
which are certainly experimentally accessible.

\section{$H \parallel$ \lowercase{$c$}}

\noindent
We first consider the field parallel to the $c$-axis.  For this
situation, all four gap nodes contribute identically. Also, we can use
symmetry to restrict the range of $\beta$ and thereby dispense with
the absolute value signs around $V$. Using the nodal approximation to
do the $\phi$ integral, we find
\bea  \label{stuff}
{N_0^{(+-)}\over\overline{N}}  & \approx &
{2\over\pi}\int_0^1\d\rho\int_0^{\pi/2}\d\beta\;P(\nu\sin\beta-\sigma\rho)
\nonumber\\
{N_0^{(--)}\over\overline{N}} & \approx &
{2\over\pi}\int_0^1\d\rho\int_0^{\pi/2}\d\beta\;P(\nu\sin\beta+\sigma\rho)
\nonumber\\
{N_0^{(-+)}\over\overline{N}} & \approx &
{2\over\pi}\int_0^1\d\rho\int_0^{\pi/2}\d\beta\;P(\sigma\rho-\nu\sin\beta).
\eea
(Recall that $P$ limits the integration ranges so that the
corresponding zero solution of the original delta function argument
exists.) It is interesting to remark that there is no limitation on
the integration range of $N_0^{(--)}$, which contributes throughout
the vortex unit cell for all values of $\sigma$ and $\nu$. The other
two terms are restricted and in fact, their allowed ranges in the
$\rho-\beta$ space are complementary. There is a physical meaning to
this. It says that in those regions of the vortex unit cell, it is
energetically favourable for the quasi-particle to flip its spin even
though this costs it diamagnetic energy. In the limit where the
diamagnetism is dominant, we have only $N_0^{(+-)}$. In the limit
where the paramagnetism is dominant, we have only $N_0^{(-+)}$, that
is all quasi-particles are spin down. In the regime where the
diamagnetism and paramagnetism are competitive, the final result
involves a nontrivial combination of both terms. These three terms can
also be combined and expressed as
\be  \label{stuff'}
{N_0\over\overline{N}}   \approx 
{4\over\pi}\int_0^1\d\rho\int_0^{\pi/2}\d\beta\;
\left(|\nu\sin\beta+\sigma\rho| + |\sigma\rho-\nu\sin\beta|\right).
\ee

From here the analysis is simple and summing all three terms we
conclude
\bea \label{bign}
{N_0 \over \overline{N}} & \approx & \sigma + {\nu^2\over2\sigma}
\phantom{{2\over\pi}\left(\left(\sigma + {\nu^2\over2\sigma}\right)
\arcsin\left({\sigma\over\nu}\right)
+ {3\nu\over2}\sqrt{1-{\sigma^2\over\nu^2}}\right)} 
\nu\leq\sigma \nonumber\\ 
& \approx & {2\over\pi}\left(\left(\sigma + {\nu^2\over2\sigma}\right)
\arcsin\left({\sigma\over\nu}\right)
+ {3\nu\over2}\sqrt{1-{\sigma^2\over\nu^2}}\right)
\phantom{\sigma + {\nu^2\over2\sigma}} \nu>\sigma.
\eea
Clearly these two expressions agree when $\sigma=\nu$. They also give
the correct behaviour for either $\sigma=0$ or $\nu=0$. For later
purposes, it is be convenient to collectively express these as
\be \label{morestuff}
{N_0\over\overline{N}} \approx F(\sigma,\nu).
\ee
The function $F$ has a nice scaling property. Specifically,
\be \label{mytown}
F(\sigma,\nu) = \nu f(\sigma/\nu)
\ee
where
\bea
f(x) & = & x + {1\over 2x}
\phantom{{2\over\pi}\left[\left(x + {1\over 2x}\right)\arcsin(x) +
{3\over2}\sqrt{1-x^2}\right]} x>1 \nonumber\\
     & = & {2\over\pi}\left[\left(x + {1\over 2x}\right)\arcsin(x) +
{3\over2}\sqrt{1-x^2}\right] \phantom{x + {1\over 2x}} x<1.
\eea
Other than the prefactor in (\ref{mytown}), the function $F$ is given
by a simple universal form.

\begin{figure}[h]
\hspace*{0.0in}\psfig{figure=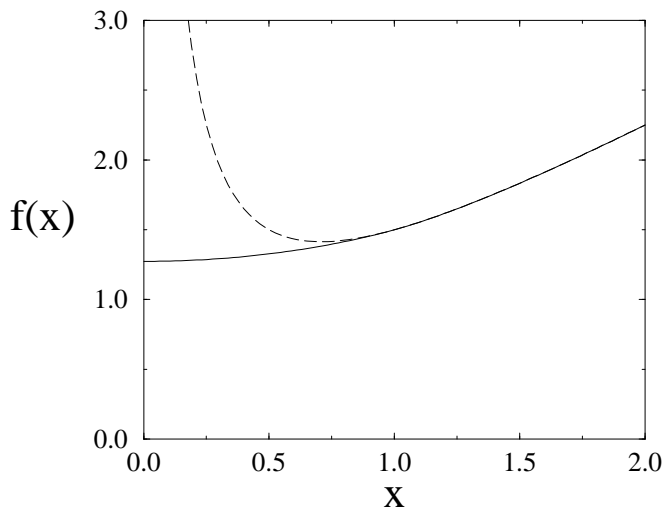,height=2.8in}
\caption
{\small The function $f$ as a function of $x$. There is a change in
functional form at $x=1$; we have continued the $x>1$ curve down to
$x=0$ as a dashed line for the sake of comparison.}
\label{hparlc}
\end{figure}

In Fig.~\ref{hparlc} we show the function $f$ plotted as a function of
$x$. To aid comparison, we plot the $x>1$ expression as a dashed line
in the range $x<1$. The change in functional form at $x=1$ represents
a very mild nonanalyticity which is not observable by eye.
Analysis of the function $f(x)$ shows that the first three derivatives
are continuous at $x=1$ and that the $x<1$ branch suffers a
$(1-x)^{7/2}$ discontinuity as it approaches $x=1$ from below.

\section{$H \parallel$ \lowercase{$ab$} and Anisotropy}

\noindent
We now consider the more interesting situation where the magnetic
field is parallel to the $ab$ planes. We continue to use the nodal
approximation and again assume that the $\phi$ integral is dominated
by the regions around the four gap nodes. We define
$\phi_n=(2n-1)\pi/4$ such that $\Delta(\phi_n)=0$ and divide the
$\phi$ integral into four domains centred on these values. In contrast
to the previous case, there is a more complicated $\phi$ dependence
since it appears in $V$ itself. However, to the same order of
approximation, we ignore the variation of $\phi$ when evaluating $V$,
and simply use the value of $\phi_n$ appropriate to that domain. The
development then follows the previous case but where we must consider
each of the four $\phi$ integrals (as indexed by $n$) and also keep
track of the $\alpha$ dependence. We also restrict the $\beta$ range
as before. The result is
\bea
{N_0^{(+-)}\over\overline{N}}  & \approx & \sum_n
{1\over2\pi}\int_0^1\d\rho\int_0^{\pi/2}\d\beta\;
P(\nu|\sin(\phi_n-\alpha)|\sin\beta-\sigma\rho) \nonumber\\
{N_0^{(--)}\over\overline{N}} & \approx & \sum_n
{1\over2\pi}\int_0^1\d\rho\int_0^{\pi/2}\d\beta\;
(\nu|\sin(\phi_n-\alpha)|\sin\beta+\sigma\rho)\nonumber\\
{N_0^{(-+)}\over\overline{N}} & \approx & \sum_n
{1\over2\pi}\int_0^1\d\rho\int_0^{\pi/2}\d\beta\;
(-\nu|\sin(\phi_n-\alpha)|\sin\beta+\sigma\rho).
\eea
By inspection each of these integrals has the same structure as
Eq.~(\ref{stuff}) but with a rescaled value for $\nu$. We then
conclude
\be
{N_0\over\overline{N}} \approx {1\over 4}\sum_n 
F(\sigma,\nu|\sin(\phi_n-\alpha)|).
\ee
By symmetry, the $n=1$ term equals the $n=3$ term and the $n=2$ term
equals the $n=4$ term.

\begin{figure}[h]
\hspace*{0.0in}\psfig{figure=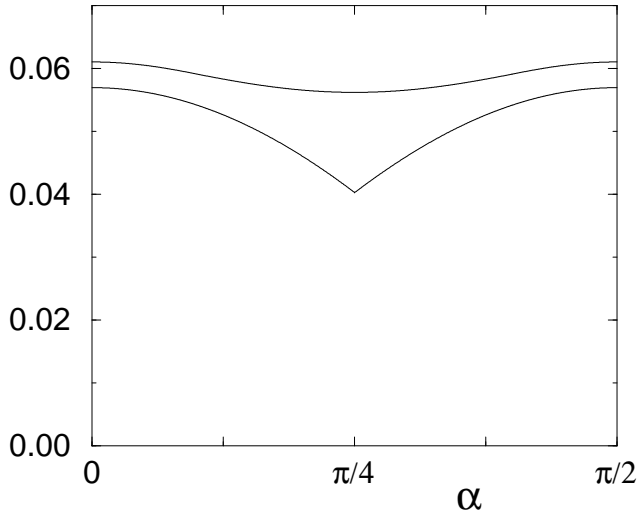,height=2.8in}
\caption
{\small The electronic density of states as a function of $\alpha$ for
$H=10T$ and using the experimentally determined value $a=0.02$. The
upper curve is for the physical value $b=0.0029$. The lower curve is
for comparison and is without paramagnetism ({\it i.e.} $b=0$.)}
\label{angdep}
\end{figure}

As mentioned, one aspect of $H\parallel ab$ is that there can be an
interesting anisotropy as we vary the angle $\alpha$. The maximum
value is for $\alpha=0$ while the minimum is for $\alpha=\pi/4$:
\bea
\left.{N_0\over\overline{N}}\right|_{\alpha=0}^{\phantom{\alpha=\pi/4}} 
& \approx & F(\sigma,\nu/\sqrt{2}) \nonumber\\ 
\left.{N_0\over\overline{N}}\right|_{\alpha=\pi/4}^{\phantom{\alpha=0}}
& \approx & {1\over 2} \left(F(\sigma,0)+F(\sigma,\nu)\right)
\eea
We define $\Delta N$ to be the difference between these two
values. There are two interesting limits that we can evaluate
analytically:
\bea \label{deltan}
\Delta N & \approx & {2\over\pi}(\sqrt{2}-1)\nu - {\sigma\over 2}
\phantom{0}\phantom{zzz} \sigma\ll\nu \nonumber\\
& \approx & 0 \phantom{{2\over\pi}(\sqrt{2}-1)\nu - {\sigma\over 2}} 
\phantom{zzz} \sigma\ge\nu.
\eea
When $\sigma=0$, the first of these is the result found in
\cite{vekhter}. We note that introducing the paramagnetism lessens the 
amount of anisotropy. This trend continues as $\sigma$ increases
until, at the critical field where $\sigma=\nu$, the anisotropy
completely vanishes. This last result can be found with more direct
physical arguments which we present in Section~\ref{heuristic}.

A relevant field scale is given by where $\Delta N$ is a
maximum. Analysis of (\ref{deltan}) shows that this happens at
$H_m=0.07H_e$. (Note that at this field, $\sigma=0.26\nu$ so that the
criterion $\sigma\ll\nu$ in (\ref{deltan}) is roughly satisfied.)  For
the theoretical estimate of $a=0.035$, we have $H_m=10$ Tesla while
for the experimental value (and the one we trust more) of $a=0.02$, we
have $H_m=3.3$ Tesla. It is this field scale which effectively
controls when the overall behaviour becomes significantly affected by
the paramagnetism.

In Fig.~\ref{angdep} we show the angular dependence of the density of
states for a fixed magnetic field as we increase the amount of
paramagnetism. As we increase the paramagnetism, the density of states
increases; this effect is particularly pronounced at $\alpha=\pi/4$,
as we expect. Consequently, the amount of anisotropy decreases, again
in agreement with our general discussion.

\begin{figure}[h]
\hspace*{0.0in}\psfig{figure=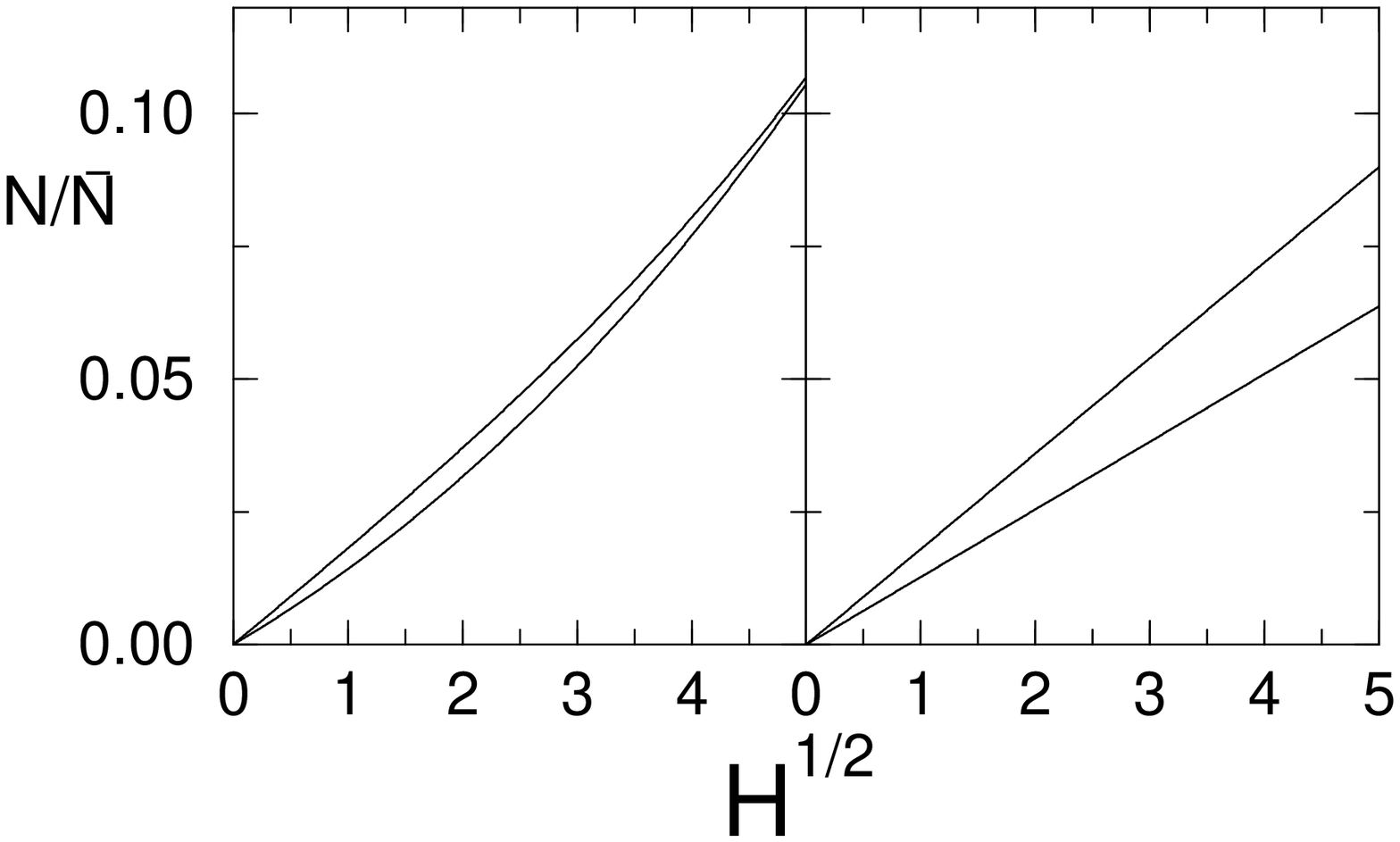,height=2.8in}
\caption
{\small The electronic density of states as a function of $\sqrt{H}$
(with $H$ measure in Tesla) for $a=0.020$. The upper curve in each
case is for $\alpha=0$ while the lower curve is for $\alpha=\pi/4$;
the difference between the curves is the magnitude of anisotropy. In
the left figure we took $b=0.0029$, the right figure is for
comparison without paramagnetism  ({\it i.e.} $b=0$.)}
\label{flddep}
\end{figure}

Next we show the anisotropy as a function of magnetic field. In
Fig.~\ref{flddep}, we plot the field dependence of the density of
states using the experimental estimates of $a=0.02$ and $b=0.0029$.
The upper curve for each pair is for the field orientation $\alpha=0$
while the lower one is for $\alpha=\pi/4$.  Notice that they start
with different slope but ultimately converge leading to a limited
range of field strengths over which the anisotropy is present. For
comparison we also show the results without anisotropy for which the
two curves never converge. Also note that around the scale $E_m$
defined above, the two curves begin to converge. At around that scale
there begins a noticeable nonlinear dependence of the density of
states on $\sqrt{H}$, which is in principle measurable. The
calculation presented here is incompatible with a density of states
which is both isotropic and linear in $\sqrt{H}$, thus providing a
clear experimental signature. The data of \cite{junod} is marginal in
that the error bars are large enough to not be in conflict with
theory. However, a more precise measurement or one that went to larger 
fields would presumably be able to settle this question. Preliminary
data from another group \cite{salamon} does seem to be consistent with 
a small anisotropy.

\begin{figure}[h]
\hspace*{0.0in}\psfig{figure=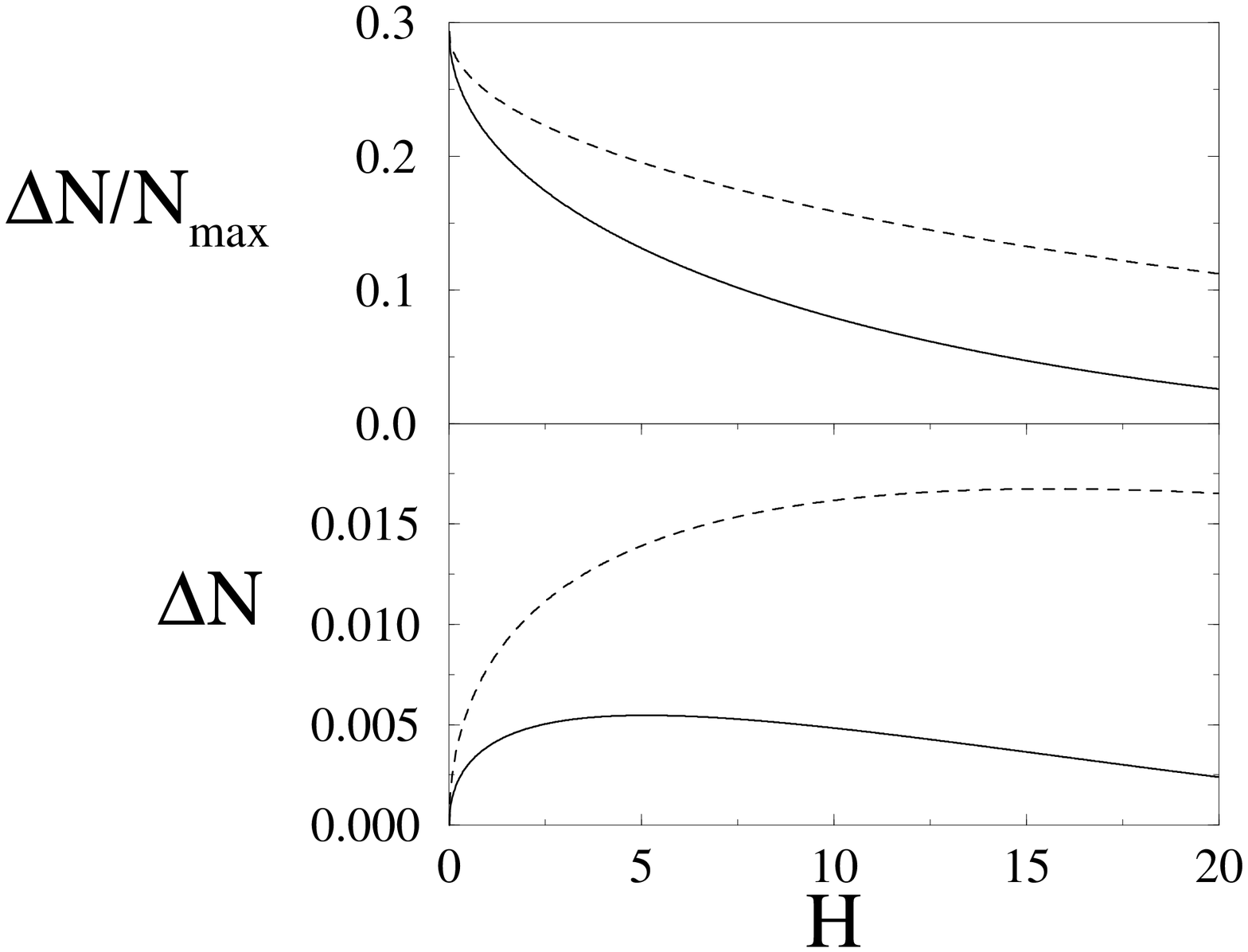,height=2.8in}
\caption
{\small The lower figure shows the absolute value of the anisotropy while
the upper figure shows the relative value of the anisotropy ($\Delta
N$ divided by the value at $\alpha=0$). The solid line is from using
$a=0.020$ while the dashed line is from using $a=0.035$.}
\label{aniso}
\end{figure}

In Fig.~\ref{aniso} we plot both the absolute anisotropy which
initially grows with magnetic field as the overall scale of the
diamagnetism increases with magnetic field. However at some
field strength of order $H_m$ it stops growing as the paramagnetic
effects begin taking hold and finally vanishes when the paramagnetic
effect dominates. We also plot the relative anistropy, {\it i.e.} the
absolute anisotropy divided by the value at $\alpha=0$. This decreases
abruptly with magnetic field (with a square root dependence for small
field) so that even relatively small values of $\sigma$ can have a
large affect on the anisotropy. We plot the results for two values of
$a$ showing the strong dependence of this affect on the diamagnetic
energy scale. Unfortunately this is not a well established number,
however we have the greatest confidence in $a=0.02$ which is shown as
the solid curve in the figure.

\section{Heuristic argument for loss of anisotropy}\label{heuristic}

\noindent
The anisotropy arises entirely from the diamagnetism, so it certainly
makes sense that it is negligible in the limit that the paramagnetic
effect is much stronger than the diamagnetic effect. However, it is
not entirely intuitive that it should disappear completely when the
effects are of comparable importance ({\it i.e.} when $\nu=\sigma$.)
In this section we present a heuristic explanation of the
anisotropy. We begin by showing a picture of what happens when a small
amount of paramagnetism is introduced to an otherwise purely
diamagnetic problem. We then present the opposite limit of near
perfect paramagnetism with a small amount of diamagnetism. We then
bridge these two limits and in the process present a concrete
explanation of why the anisotropy switches off precisely when
$\nu=\sigma$.

\begin{figure}[h]
\hspace*{0.0in}\psfig{figure=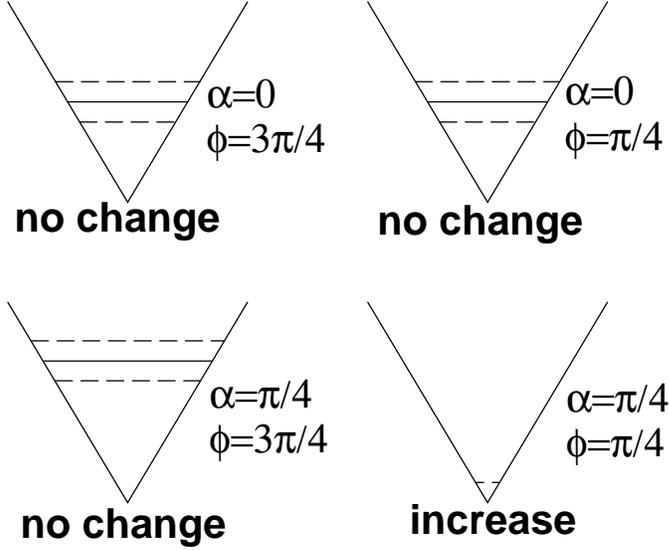,height=2.8in}
\vspace{.5cm}
\caption
{\small Each ``V'' shows the dependence of the density of states on
the argument $\omega$. It is locally linear but with a cusp at zero
argument. We show the results for the field aligned along the gap
antinode (top two) and aligned along the gap node (bottom two). In
each row, the left hand figure is for the $\phi=3\pi/4$ gap node and
the right hand figure is for the $\phi=\pi/4$ gap node. The solid line
in each case shows the result with no paramagnetism (note that there
is no contribution from the bottom right diagram in this case.) The
contribution from the two nodes for $\alpha=0$ outweighs the one
contribution for $\alpha=\pi/4$. We indicate the effect of adding
paramagnetism with the dashed lines. Each solid line gets split into
two contributions but the average of the two equals the result without
paramagnetism, due to the linear behaviour, so there is no change to
the net contribution. For the bottom right diagram this is not true
and there is a net increase in the contribution. This lessens the
anisotropy.}
\label{new}
\end{figure}

In Fig.~\ref{new} we show the dependence of the density of states as a
function of frequency, in brief $N(\omega)\propto|\omega|$. Although
all of our calculations involve the physical frequency being zero, due
to the Doppler and paramagnetic shifts, our final answers involve
evaluating this density of states at various arguments. For example,
in Eq.~(\ref{stuff'}), we can interpret the integrands
$|\sigma\rho\pm\nu\sin(\phi-\alpha)|$ as being the density of states
$N(\omega)$ evaluated at $\omega=\sigma\rho\pm\nu\sin(\phi-\alpha)$.
All of the interplay between the paramagnetism and diamagnetism
involves arguments being near zero so that the cusp in the function
plays an important role. For concreteness, we take $\beta=\pi/2$ and
$\rho=1$; we will discuss what happens elsewhere in the vortex unit
cell in the subsequent discussion. In the purely diamagnetic limit, we
set $\sigma=0$. For $H\parallel c$, all nodes contribute identically
and there is nothing interesting to say. Instead we will focus on
$H\parallel ab$. In this case, the nodes at $\phi=\pi/4$ and $5\pi/4$
are identical as are the nodes at $\phi=3\pi/4$ and $7\pi/4$, so we
will only look at one of each of these pairs. Recall that the rescaled
Doppler shift is $|\nu\sin(\phi-\alpha)|$. When $\alpha=0$, all nodes
contribute identically as $\nu/\sqrt{2}$; as indicated in the top two
diagrams of the figure. When $\alpha=\pi/4$, one pair of nodes
contributes as $\nu$ while the other does not. The dependence on the
argument is linear, meaning a greater density of states for $\alpha=0$
than for $\alpha=\pi/4$ (since $2\nu/\sqrt{2}>\nu$.)

Imagine now that we add a small amount of paramagnetism, in other
words set $\sigma$ to be a small but non-zero value. This induces a
splitting in the argument at which we evaluate $N(\omega)$ to
$\nu|\sin(\phi-\alpha)|\pm\sigma$. As long as $\nu\sin(\phi-\alpha)$
is larger in magnitude than $\sigma$, this splitting has no
effect. This is because $N(\omega)$ is locally linear so that the
extra amount from one sign of $\sigma$ is exactly compensated for by
the other sign of $\sigma$. The only case that this does not apply to
is the $\phi=\pi/4$ node when $\alpha=\pi/4$. In this event, both
signs of $\sigma$ contribute equally and there is no cancellation.
This implies an increase in the density of states for $\alpha=\pi/4$
and no change for $\alpha=0$ so that the anisotropy is lessened, in
accord with Eq.~(\ref{deltan}). (We recall that this was all for the
one point $\rho=1$ and $\beta=\pi/2$. When we average over the vortex
unit cell, there are regions of small $\beta$ where $\sigma\rho$
dominates the diamagnetic shift. This leads to corrections of order
$\sigma^2$ to the final, average density of states. This does not
affect our general conclusion that adding paramagnetism lessens the
anisotropy.)

We now consider the opposite limit where the paramagnetism
dominates. With no diamagnetism, we simply evaluate the density of
states at $\pm\sigma$. This is clearly the same for all orientations
and there is no anisotropy. We now add a small amount of diamagnetism,
in other words set $\nu$ to be a small but non-zero value. Analogously
to before, we have shifts in the argument by the amount
$\pm\nu\sin(\phi-\alpha)$. As before, the local linear dependence of
$N(\omega)$ means that the net effect cancels and there is no change
to the densities of states and hence no introduction of anisotropy.
Again, recall that this was for $\beta=\pi/2$ and $\rho=1$. The
argument just presented actually breaks down for small $\rho$ where
the diamagnetism is greater than the paramagnetism and for which an
altogether different mechanism preserves the lack of anisotropy. The
diamagnetism is dominant in the region of the unit cell for which
$\sigma\rho<\nu|\sin(\phi-\alpha)\sin\beta|$; the area of this
region clearly scales as $\nu$.  The value of the density of states in
this region scales as $\nu$ (due to the local linear dependence of
$N(\omega)$). Therefore, the total contribution scales as
$\nu^2$. This is small when the paramagnetism dominates.  Furthermore
it leads to no anisotropy since the resultant sum over the nodes is
proportional to $\sin^2(\pi/4-\alpha)+\sin^2(3\pi/4-\alpha)=1$, which
is clearly independent of $\alpha$. The key points are the quadratic
scaling in $\nu$ and the rescaling of $\nu$ by $\sin(\phi-\alpha)$.
Taken together they lead to a completely isotropic $\alpha$
dependence.

For cases where $\sigma$ and $\nu$ are comparable, we continue to use
the arguments of the above paragraph. As long as $\nu<\sigma$, the
scaling with $\nu^2$ is perfect (as is also apparent in
Eq.~(\ref{bign})) and there is no anisotropy. As soon as $\nu>\sigma$,
the scaling breaks down. This is because the curve
$\sigma\rho=|\nu\sin(\phi-\alpha)\sin\beta|$ exceeds the maximum value
of $\rho=1$ and the area of the vortex unit cell in which the
diamagnetism dominates begins to saturate. The contribution of this
part of the vortex unit cell then increases more slowly than $\nu^2$
and a nontrivial $\alpha$ dependence becomes possible. Nevertheless,
the scaling is still approximately maintained even for $\nu>\sigma$ so
that the anisotropy turns on very slowly. (This is related to the very
smooth behaviour of the function $f(x)$ near $x=1$ shown in
Fig.~\ref{hparlc}.) Thus even for $\sigma$ quite a bit smaller than
$\nu$, the anisotropy is strongly suppressed.

We remark that we made extensive use of the local linear behaviour of
the density of states. This linear behaviour arises from using the
nodal approximation for small fields. A more general discussion
allowing for quadratic and higher powers will not affect the general
conclusions but presumably would lead to a small amount of anisotropy
for $\sigma>\nu$.

\section{Conclusion}

\noindent
In this paper we have discussed the combined effect of paramagnetism
and diamagnetism on the zero frequency quasiparticle density of states
in the presence of a magnetic field. When the field is parallel to the
$c$-axis, there is no interesting anisotropic effect. Furthermore, for
the cuprates, the energy scales are such that the paramagnetism is
negligible for any physically realisable magnetic field strength.

When the field is in the $ab$ planes, the diamagnetism is less
important --- this is related to the fact that it is more difficult to
form vortices out of the copper-oxide planes. In fact, we expect that
it is the paramagnetism which is dominant for all materials but
YBCO. YBCO is special since the $c-$axis transport, necessary for the
formation of vortices out of the plane, is small but not negligeable.
For this material, we expect that the paramagnetic and diamagnetic
effects to be of comparable importance for magnetic fields of
experimental interest. One important issue is the anisotropy as we
change the angle between the applied field direction and the gap
nodes. Even a small amount of paramagnetism strongly suppresses the
expected anisotropic response, providing a possible explanation of why
this effect has not been observed in experiments. However, the same
field scale where the paramagnetism begins to strongly suppress the
anisotropy also controls where the linear dependence of $N$ on
$\sqrt{H}$ begins to fail, as is apparent in Fig.~\ref{flddep}.
Therefore, if this is the explanation, one should also observe
nonlinearities in such experimental curves. The experiments of
\cite{junod} seem to indicate both no anisotropy and linear behaviour in
$\sqrt{H}$, in contradiction with our conclusions. However, the error
bars are large enough to still be consistent with our conclusions.
More precise experiments or ones using larger fields are necessary to
resolve this question. Preliminary data from \cite{salamon} indicates
that there is an observable anisotropy.

Research supported in part by the Natural Sciences and Engineering
Research Council (NSERC) and by the Canadian Institute for Advanced
Research (CIAR).

\appendix

\section{General Results for $H\parallel$ \lowercase{$c$}}

\noindent
In this appendix we generalise the results found using the nodal
approximation by not considering either $\sigma$ or $\nu$ to be small
but limiting ourselves to the somewhat simpler $H\parallel c$ case.
Consider equations (\ref{general}). For a given choice of $\rho$ and
$\beta$, there is always some finite range of $\phi$ such that a
solution to the $(--)$ equation exists; this is not true for the
$(+-)$ and $(-+)$ equations. The resulting $\phi$ integrals correspond
to the definition of the elliptic function and we conclude
\bea \label{n0terms}
{N_0^{(--)}\over \overline{N}} & = & {4\over\pi^2}\int\d\beta\d\rho\rho\;
G(a)\nonumber\\
{N_0^{(-+)}\over \overline{N}} & = & {4\over\pi^2}\int_{V<\mu H}\d\beta\d\rho\rho\;
G(b)\nonumber\\
{N_0^{(+-)}\over \overline{N}} & = & {4\over\pi^2}\int_{V>\mu H}\d\beta\d\rho\rho\;
G(c)
\eea
where we have defined the function
\bea
G(x) & = & K(x) 
\phantom{{1\over x} K\left({1\over x}\right)} x<1 \nonumber\\
                   & = & {1\over x} K\left({1\over x}\right)
\phantom{K(x)} x\ge 1.
\eea
$K(x)$ is the complete elliptic integral of the first kind and the
parameters are
\be
a = {\rho \over \nu\sin\beta + \sigma\rho}
\hspace{.7cm}
b = {\rho \over \sigma\rho - \nu\sin\beta}
\hspace{.7cm}
c = {\rho \over \nu\sin\beta-\sigma\rho}.
\ee
There is a logarithmic singularity when the argument of $G$ equals
unity but this is integrable. The integration domain of the first
equation in (\ref{n0terms}) is $0<\rho\leq 1$ and $0<\beta\leq\pi/2$
so that $V=|V|>0$. The domain is the same for the other two except for 
the added constraint on $V-\mu H$.

It is difficult to go further in an explicit analytic determination of
the integrals. However, we can say a few things on a qualitative
level. There are logarithmic singularities whenever one
of the quantities $a$, $b$ or $c$ is unity. These possibilities
correspond to curves of singularities in the $\rho-\beta$ space. Also,
when $b$ (or, equivalently, $c$) equals infinity, there is a curve of
zeros in the $\rho-\beta$ space. As $\nu$ and $\sigma$ are varied,
singular things can happen to these special curves. For example, in
order to have a logarithmic singularity in $N_0^{(--)}$, we require
that $a=1$ which in turn requires
\be
\rho={\nu\over1-\sigma}\sin\beta.
\ee
Clearly if $\sigma>1$ there is no solution to this condition so that
as $\sigma$ crosses through unity we will notice the abrupt
disappearance of the curve of logarithmic singularities in the
integrand of $N_0^{(--)}$. For $N_0^{(-+)}$, it is precisely the
converse. There are only logarithmic singularities if $\sigma>1$ and
none if $\sigma<1$. Therefore, as $\sigma$ crosses through unity, we
expect some nonanalyticity in the density of states.

Another possibility for $N_0^{(--)}$ is that the curve of
singularities crosses through $\rho=1$ and out of the allowed domain
of the $\rho$ integral. By the previous equation, this will happen
when $\nu=1-\sigma$. The analogous event happens for $N_0^{(-+)}$ when
$\nu=\sigma-1$ and for $N_0^{(+-)}$ when $\nu=1+\sigma$. The final
possibility is that the curve of zero values for $N_0^{(+-)}$ and
$N_0^{(-+)}$ crosses through $\rho=1$ and out of the allowed domain of
the $\rho$ integral. The curve of zero values (when $b$ and $c$ are
infinite) requires $\rho=\nu\sin\beta/\sigma$ so this event occurs
when $\nu=\sigma$. We already encountered the $\sigma=\nu$ criterion
in our small argument analysis and is shown in Fig.~1. The others all
require that either $\nu$ or $\sigma$ (or both) be of the order of
unity and so were not captured in our small argument analysis.

\begin{figure}[h]
\hspace*{0.0in}\psfig{figure=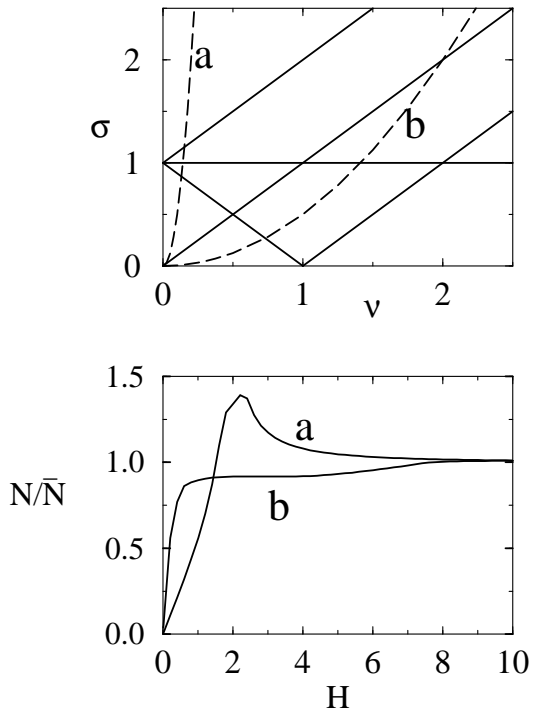,height=3.8in}
\caption
{\small Top: A plot of the different lines in the $\nu-\sigma$ space
along which we expect nonanalyticities in the density of states.  The
two parabolas represent possible traversals through this space as we
increase the magnetic field. Bottom: the resultant dependence of the
density of states for the two parabolas of the upper plot. The
horizontal axis is in arbitrary units.}
\label{singularities}
\end{figure}

A plot of the different domains in the $\nu-\sigma$ space is shown in
Fig.~\ref{singularities}. Each of the solid lines represents some
border along which one of the three terms experiences a
nonanalyticity. As we change the magnetic field, we follow a parabola
in this parameter space and thereby intersect various of these lines.
The precise sequence of intersections depends on the energy scales
associated with the paramagnetic and diamagnetic terms. There are five
possible sequences depending on the sharpness of the parabola; we show
two of them. For a very sharp parabola, the paramagnetic energy is
dominant except at very small fields and the result is similar to a
purely paramagnetic calculation. The major differences are 1) an
initial $\sqrt{H}$ dependence for small enough fields and 2) the
logarithmic singularity at $\sigma=1$, which would be present for pure
paramagnetism, is softened by the diamagnetism.

In a situation where the diamagnetism is dominant for low fields, the
magnetic field dependence is rather similar to a purely diamagnetic
case in which the density of states rises smoothly and saturates at
some critical value (at $\nu=1$.) Again, this is softened, there is
additional structure once the paramagnetism becomes comparable to the
diamagnetism ({\it i.e.} $\sigma\sim\nu$.) For large enough fields the
paramagnetism is always dominant. This means that $N/\overline{N}$ will
always approach unity from above.

Unfortunately, most of this discussion is academic. As argued above,
the cross-over field where $\sigma=\nu$ for $H\parallel c$ is about
1000T. The value where $\sigma=1$ is about 350T while the value where
$\nu=1$ is about 70T. These are the scales where these affects would
be visible, and they are probably beyond any experimental reach. For
$H\parallel ab$ it is reasonable to expect a similar structure to that
shown in Fig.~\ref{singularities}. The field where $\sigma=\nu$ is
about 100T. The value where $\sigma=1$ is still 350T and the value
where $\nu=1$ is about 1000T (again, depending on the value chosen for
$a$). Again, all of these fields are beyond experimental reach and so
the results of this appendix could probably not be observed in
experiment. Nevertheless it is of theoretical interest to understand
the analytic structure of the density of states.

\end{document}